\documentstyle[twoside,wicsbook,graphicx]{article}

\begin{document}

\title{\vspace{-3pc}\titlesize A Neural Network-based ARX Model of Virgo Noise.}
\author{{\large F. Barone, R. De Rosa, A. Eleuteri, F. Garufi, L. Milano} \\
{\normalsize Dipartimento di Scienze Fisiche, Universit\`{a} di Napoli
''Federico II'' }\\
{\normalsize Istituto Nazionale di Fisica Nucleare, sez. Napoli, }\\
{\normalsize Complesso Universitario di Monte Sant'Angelo, }\\
{\normalsize via Cintia, I-80126 Napoli Italia}\\
\and {\large R. Tagliaferri} \\
{\normalsize Dipartimento di Matematica ed Informatica, }\\
{\normalsize Universit\`{a} di Salerno, and INFM unit\`{a} di Salerno,}\\
{\normalsize via S. Allende, 84081 Baronissi (SA) Italia }\\
{\normalsize and IIASS ''E. R. Caianiello'', Vietri s/m Italia}}
\maketitle
\thispagestyle{empty}
\begin{abstract}
\ninesize
\noindent In this paper a Neural Networks based approach is presented to
identify the noise in the VIRGO context. VIRGO is an experiment to detect
Gravitational Waves by means of a Laser Interferometer. Preliminary results
appear to be very promising for data analysis of realistic Interferometer
outputs.
\end{abstract}



\section{Introduction}

Neural Networks (NN's) have become in the last years a very effective
instrument for solving many difficult problems in the field of Signal
Processing due to their properties like non-linear dynamics, adaptability,
self-organization and high speed computational capability (see for example 
\cite{Luo97} and the papers therein quoted).

Aim of this paper is to show the feasibility of the use of NN's to solve
difficult problems of signal processing regarding the so called VIRGO
project. Gravitational Waves (GW's) are travelling perturbations of the
space-time predicted by the theory of \ General Relativity, emitted when
massive systems are accelerated. Up to now, there is only an indirect
evidence of their existence, obtained by the observations of the binary
pulsar system PSR 1913+16. Moreover, the direct detection of GW's is not
only a relevant test of General Relativity, but the start of a new picture
of the Universe. In fact, GW's carry complementary information with respect
to electromagnetic and optical waves, since the GW's are practically not
absorbed by the matter.

The aim of the Virgo experiment is the direct detection of gravitational
waves and, in joint operation with other similar detectors, to perform
gravitational waves astronomical observations. In particular, the VIRGO
project is designed for broadband detection from $10Hz$ to $10kHz$. The
principle of the detector is shown in figure~1.

 A $3$ $km$ arm-length Michelson
interferometer with suspended mirrors (test masses) is used. The phase
difference $\Delta \phi $ between the two arms is amplified using
Fabry-Perot cavities of Finesse $50$ in each arm. Aiming for detection
sensitivity of $3.10E-23\frac{1}{Hz}\quad at\quad 100Hz$, VIRGO is a very
delicate experimental challenge because of the competition between various
sources of noise and the very small expected signal. In fact, the
interferometer will be tuned on the dark fringe, and then the signal to
noise ratio will be mainly limited, in the above defined range of
sensitivity, by residual seismic noise, thermal noise of the suspensions
photon counting noise (shot noise). In figure 2 the overall sensitivity of
the apparatus is shown. 
In this figure it is easy to see the contribution of
the different noise sources to the global noise.

In this context we use a Multi-Layer Perceptron (MLP) NN with
the back-propagation learning algorithm to model and identify the noise in the
system, because we experimentally found that FIR NN's and Elman NN's did not work in a satisfying manner.

Both the FIR \cite{Tsoi91} and Elman \cite{Elman90} models proved to be very
sensible to overfitting and were not stable.  Furthermore the Elman network
required a great number of hidden units, while the FIR network required a
great number of delay terms. Instead, the MLP proved succesfull and easy to
train because we used the Bayesian learning paradigm.

\section{Neural networks for time-domain system identification}

NN's are massively parallel, distributed processing systems. They
are composed of a large number of processing elements (called nodes, units
or neurons) which operate in parallel. Scalars (called weights) are
associated to the connections between units and determine the strength of
the connections themselves. Computational capability is due to the
connections between the units and to their collective behaviour.
Furthermore, information representation is distributed, i.e. no single unit
is charged with specific information.
NN's are well-known for their universal approximation capability
\cite{Hornik89}.
 
System identification consists in finding the input-output mapping of a
dynamic system. A discrete-time Multi-Input Multi-Output (MIMO) system (or a
continuous-time sampled-data system) can be described with a set of
non-linear difference equations of the form ({\em input-state-output}
representation): 
\begin{equation}
\left\{ 
\begin{array}{ll}
{\bf z}(n+1)=F({\bf z}(n),{\bf u}(n)) &  \\ 
{\bf y}(n)=G({\bf z}(n)) & 
\end{array}
\right.  \label{eq:iso}
\end{equation}
where ${\bf z}\in R^{L}$ is the state vector of the system, ${\bf u}\in
R^{M} $ is the input vector and ${\bf y}\in R^{N}$ is the output vector. 
Since we
can not always access the state vector of the system, therefore we can use an
input-output representation of the form: 
\begin{eqnarray}
{\bf y}(n) &=&{\cal F}[{\bf y}(n-1),{\bf y}(n-2),\ldots ,{\bf y}(n-n_{y}), 
\nonumber \\
&&{\bf u}(n-n_{d}),\ldots ,{\bf u}(n-n_{u}-n_{d}+1)]  \label{eq:io}
\end{eqnarray}
where $n_{u}$ and $n_{y}$ are the maximum lags of the input and output
vectors, respectively, and $n_{d}$ is the pure time-delay (or dead-time) in
the system. This is called an ARX model (autoregressive with exogenous
inputs,) and it can be shown that a wide class of discrete-time systems can
be put in this form~\cite{Chen89}. To build a model of the form~(\ref{eq:io}%
), we must therefore obtain an estimation of the functional ${\cal F}[\cdot
] $, which generally is nonlinear.

Given a set of input-output pairs, a neural network can be built \cite{Luo97}
which approximates the desired functional ${\cal F}[\cdot ]$. Such a network
has $n_{y}N+(n_{u}-n_{d})M$ inputs and $N$ outputs (see figure 3). 

A
difficulty in this approach arises from the fact that generally we do not
have information about the model order (i.e. the maximum lags to take into
account) unless we have some insight into the system physics. Furthermore,
the system is non-linear. Recently \cite{He93} a method has been proposed for
determining the so-called {\em embedding dimension} of nonlinear dynamical
systems,
 when the input-output pairs are affected by very low noise.
Furthermore, the lags can be determined by evaluating the \emph{average mutual
information} (AMI)\cite{Abarbanel96}. Such methodologies, although not always
successful, can be nevertheless used as a starting point in model design.

\section{Virgo}
 
In the VIRGO data analysis, the most difficult problem is the gravitational
signal extraction from the noise due to the intrinsic weakness of the
gravitational waves, to the very poor signal-to-noise ratio and to their not
well known expected templates. Furthermore, the Virgo detector is not yet
operational, and the noise sources analyzed are purely theoretical models
(often stationary noises), not based on experimental data. Therefore, we
expect a great difference between the theoretical noise models and the
experimental ones. As a consequence, it is very important to study and to
test algorithms for signal extraction that are not only very good in signal
extraction from the theoretical noise, but also very adaptable to the real
operational conditions of Virgo.

For this reason, we decided the following strategy for the study, the
definition and the tests of algorithms for gravitational data analysis. The
strategy consists of the following independent research lines.

The first line starts from the definition of the expected theoretical noise
models. Then a signal is added to the Virgo noise generated and the
algorithm is used for the extraction of the signal of known and unknown
shape from this noise at different levels of signal-to-noise ratio. This
will allow us to make a number of data analysis controlled experiments to
characterize the algorithms.

The second line starts from the real measured environmental noise (acoustic,
electromagnetic, ...) and tries to identify the noise added to a theoretical
signal. In this way we can test the same algorithms in a real case when the
noise is not under control.

Using this strategy, at the end, when in a couple of years Virgo will be
ready for the first test of data analysis, the procedure will be moved to
the real system, being sure to find small differences from theory and
reality after having acquired a large experience in the field.

\section{A neural network-based model of the Virgo system}

As we have seen in the introduction, the Virgo interferometer can be
characterized by a sensitivity curve, which expresses the capability of the
system to filter undesired influences from the environment, and which could
spoil the detection of gravitational waves (such a noise is generally called
seismic noise). The sensitivity curve has the following expression: 
\begin{equation}
S(f)=\left\{ 
\begin{array}{ll}
\frac{S_{1}}{f^{5}}+\frac{S_{2}}{f}+S_{3}\left[ 1+\left( \frac{f}{f_{k}}
\right) ^{2}\right] +S_{\nu } & \quad f\geq f_{\textrm{min}} \\ 
S(f_{\textrm{min}}) & \quad f<f_{\textrm{min}}
\end{array}
\right.   \label{eq:exprsenscurve}
\end{equation}
where:

\begin{itemize}
\item  $f_{\textrm{min}}=4Hz$

\item  $f_{k}=500Hz$ is the shot noise cut-off frequency

\item  $S_{1}=1.8\cdot 10^{-36}$ is the pendulum mode

\item  $S_{2}=0.33\cdot 10^{-42}$ is the mirror mode

\item  $S_{3}=3.24\cdot 10^{-46}$ is the shot noise
\end{itemize}

The contribution $S_{\nu }(f)$ of violin resonances is given by: 
\begin{equation}
S_{\nu }(f)=\sum_{i}^{n}\frac{1}{i^{4}}\frac{f_{i}^{(c)}}{f}\frac{C_{c}\phi
_{i}^{2}}{\left[ \left( \frac{f}{if_{i}^{(c)}}\right) ^{2}-1\right]
^{2}+\phi _{i}^{2}}+\frac{1}{i^{4}}\frac{f_{i}^{(f)}}{f}\frac{C_{f}\phi
_{i}^{2}}{\left[ \left( \frac{f}{if_{i}^{(f)}}\right) ^{2}-1\right]
^{2}+\phi _{i}^{2}}  \label{eq:violinres}
\end{equation}
where the different masses of close and far mirrors are taken into account:

\begin{itemize}
\item  $f_{i}^{(c)}=i\cdot 327Hz$

\item  $f_{i}^{(f)}=i\cdot 308.6Hz$

\item  $C_{c}=3.22\cdot 10^{-40}$

\item  $C_{f}=2.82\cdot 10^{-40}$

\item  $\phi _{i}^{2}=10^{-7}$
\end{itemize}

Note that we used
a simplified curve for our simulations, in which we neglected the resonances (see
figure 4).

Samples of the sensitivity curve $\{S_{i}\}_{i}$ can be obtained by
evaluating the expression (\ref{eq:exprsenscurve}) at a set of frequencies $%
\{f_{i}\}_{i}$, $f_{i}\in \lbrack 10,10000]$. The samples of the sensitivity
curve allow us to obtain the system transfer function (in the frequency
domain), $N(j2\pi f)$, such that: 
\begin{equation}
|N(j\omega )|^{2}=S(j\omega )  \label{eq:tranfun}
\end{equation}
>From this, by means of an inverse Discrete Fourier Transform, samples of the
system transfer function (in the time domain) can be obtained. Our aim is to
build a model of the system transfer function (\ref{eq:tranfun}).

Assuming that the interferometer input noise is a zero mean Gaussian
process, by feeding it to the system (i.e. filtering it through the system
transfer function) we obtain a coloured noise. The so obtained white
noise-colored noise pairs can then be used to train an MLP, as shown in
figure 3.

\subsection{Experimental Results}

The first step in building an ARX model is the model order determination. To
determine suitable lags which describe the system dynamics, we used the
AMI criterion \cite{Abarbanel96}. 
This can be seen as a generalization of the autocorrelation function, used to
determine lags in linear systems. A strong property of the AMI statistic is
that it takes into account the non-linearities in the system. Usually, the
lag is chosen as the first minimum of the AMI function. The result is reported
in figure~5, in which the first minimum is at 1.
To find how many samples are necessary to unfold the (unknown) state-space of
the model (the so called \emph{embedding dimension} \cite{Abarbanel96}) we used
the method of \cite{He93}, the Lipschitz decomposition. The result of the
search is reported in
 figure~6.

From the figure we can see that,
starting from lag three, the order
 index decreases very slowly, and so we can
derive that the width of the input window is at
 least three. In order to test
the NN's capability in solving the problem, we
chose a width of 5, both for
input and output (i.e. $n_{y}=n_{u}=5$). In
 this way, we obtained a NN with a
simple structure. Furthermore, some
 preliminary experiments showed that the
system dead-time is $n_{d}=0$; this
 gives the best description of the system
dynamics.

Another fundamental issue is the NN complexity, i.e. the number
of units in
 the hidden layers of the NN. Usually the determination of the
network complexity is critical because of the risks of overfitting.
Since 
the NN was trained following a Bayesian framework, then overfitting was of no
concern; so we directed our search for a model with the minimum possible
complexity. In our case, we found a hidden layer with 6 $tanh$ units is
optimal.

The Bayesian learning framework (see \cite{MacKay94} and \cite{Neal94}) allows
the use of a \emph{distribution} of NN's,
that is, the model is a realization of a random vector whose
components are the NN weights. The so obtained NN is the {\em most probable}
given the data used to train it. This approach avoids the bias of the
cross-validatory techniques commonly used in practice to reduce model
overfitting \cite{Bishop96}. To allow for a smooth mapping which does not
produce overfitting, several regularization parameters (also called
\emph{hyperparameters}) embedded in the error criterion have been used:
\begin{itemize}
\item one for each set of connections out of each input node,
\item one for the set of connections from hidden to output units,
\item one for each of the bias connections,
\item one for the data-contribution to the error criterion.
\end{itemize}
Usually, the hyperparameters of the first three kinds are called \emph{alphas},
while the last is called a \emph{beta}.
 
The approach followed in the
application of the Bayesian framework is the ``exact integration'' scheme,
where we sample from the analytical form of the distribution of the network
weights. This can be done if we assume an analytic form for the prior
hyperparameters distribution, in particular a distribution uniform on a
logarithmic scale: 
\begin{displaymath}
p(\ln \alpha _i) = \frac{1}{\alpha _i}.
\end{displaymath}
This distribution is \emph{non-informative}, i.e. it embodies our complete
lack of knowledge about the values the hyperparameters should take.

The chosen model was trained using a sequence of little less than a million of
patterns (we sampled the system at 4096Hz for 240s) normalized to zero mean
and unity variance with a \emph{whitening} process. Note that the
 input-output
pairs were processed through discrete integrators to obtain
 pattern-target
pairs, as shown in figure 3. The NN was then tested on a 120s long sequence.

The NN was trained for $370$ epochs, with the hyperparameters being
updated every $15$ epochs. A close look at the $\alpha $ hyperparameters
shows that all the inputs are relevant for the model (they are of the same
magnitude; note that this further confirms the pre-processing analysis). The
$\beta $ hyperparameter shows that the data contribution to the error is very
small (as we would expect, since the data are synthetic).

The simulations were made using the MATLAB$^{\copyright}$
language, the Netlab Toolbox~\cite{Bishopetal96} and other software designed
by us.
 
In figure~7, the PSDs of the target and the predicted time series are shown; in the lower 
part of the figure is reported the PSDs of the prediction residuals.

In figure~8, the PSD of a 100Hz sine wine added to the noise is shown, with the signal 
extracted by the network. As can be seen, the network recognizes the frequency of the
sine wave with the maximum precision allowed by the residuals.

\section{Conclusion}

\bigskip

In this paper we have shown some preliminary tests on the use of NN's for
signal processing in the VIRGO Project. Some observations can be elicited
from the experimental results:

\begin{itemize}
\item  In evaluating the Power Spectral Densities (PSDs), we made the
hypothesis that the system is sampled at $4096Hz$. It is only a work
hypothesis, but it shows how the network reproduces the system dynamics up
to $2048Hz$. Note that the PSDs are nearly the same also if we were near
the Nyquist frequency.

\item  The PSD of the residuals shows a nearly-white spectrum, which is
index of the model goodness (see \cite{Ljung95}).

\end{itemize}

The next steps in the research are: 

\begin{description}
\item[-]  to increase the system model order and to test if there are
significant differences in prediction;

\item[-]  to test the models with a greater number of samples to obtain a
better estimate of the system dynamics;

\item[-]  to model the noise inside the system model to improve the system
performance and to allow a multi-step ahead prediction (i.e. an output-error
model)
\end{description}

\newpage

\begin{figure}
\begin{center}
\includegraphics[width=12cm,height=10cm]{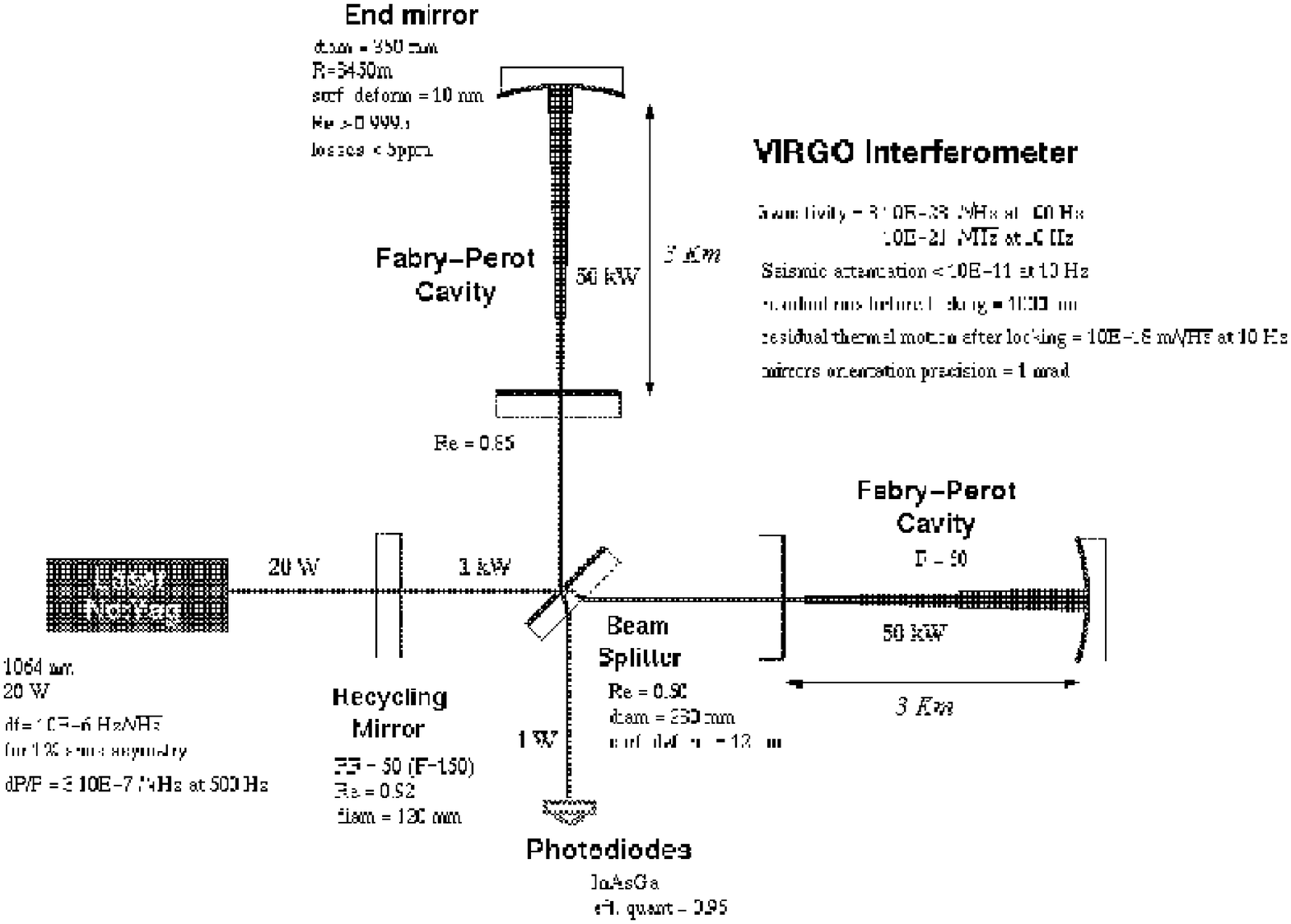}
\end{center}
\caption{Scheme of the Virgo interferometer.}
\end{figure}

\begin{figure}
\begin{center}
\includegraphics[width=11cm,height=8cm]{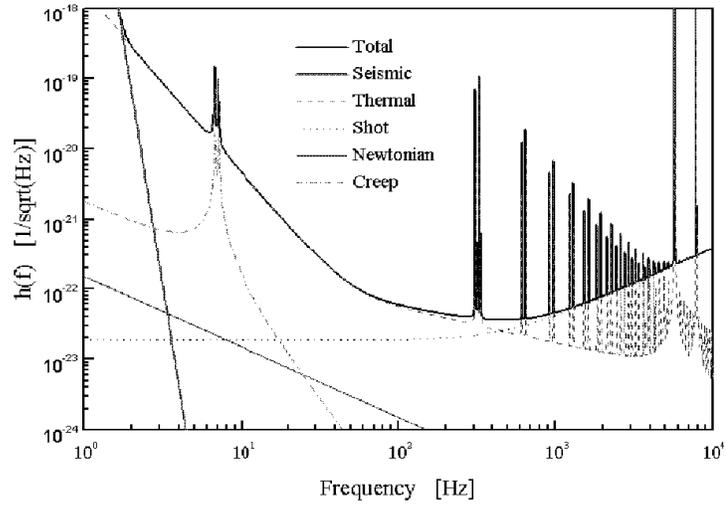}
\end{center}
\caption{Sensitivity curve of the Virgo interferometer.}
\end{figure}

\begin{figure}
\begin{center}
\includegraphics[width=10cm,height=8cm]{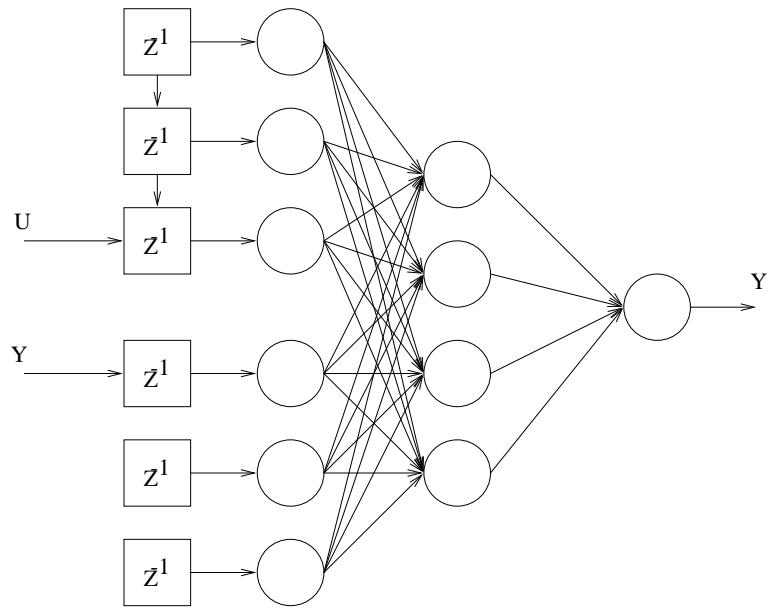}
\end{center}
\caption{A neural network ARX model.}
\end{figure}

\begin{figure}
\begin{center}
\includegraphics[width=10cm,height=8cm]{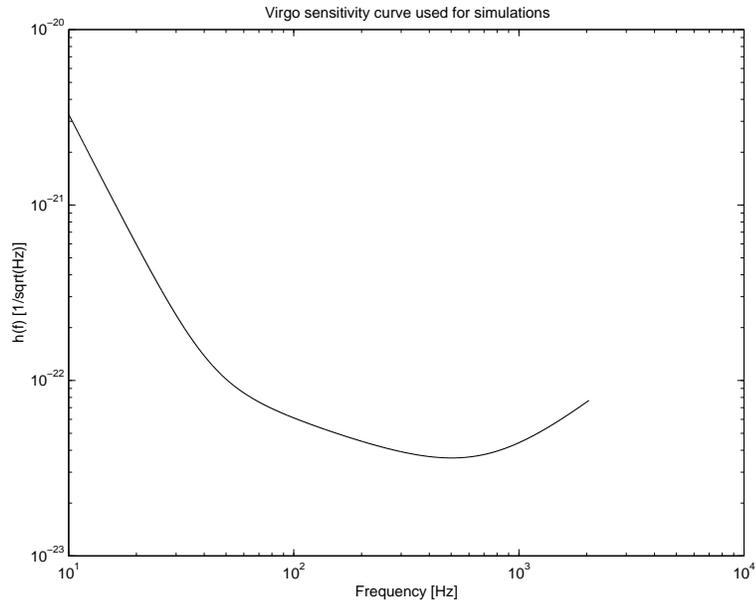}
\end{center}
\caption{The simplified sensitivity curve used for the experiments.}
\end{figure}

\begin{figure}
\begin{center}
\includegraphics[width=11cm,height=8cm]{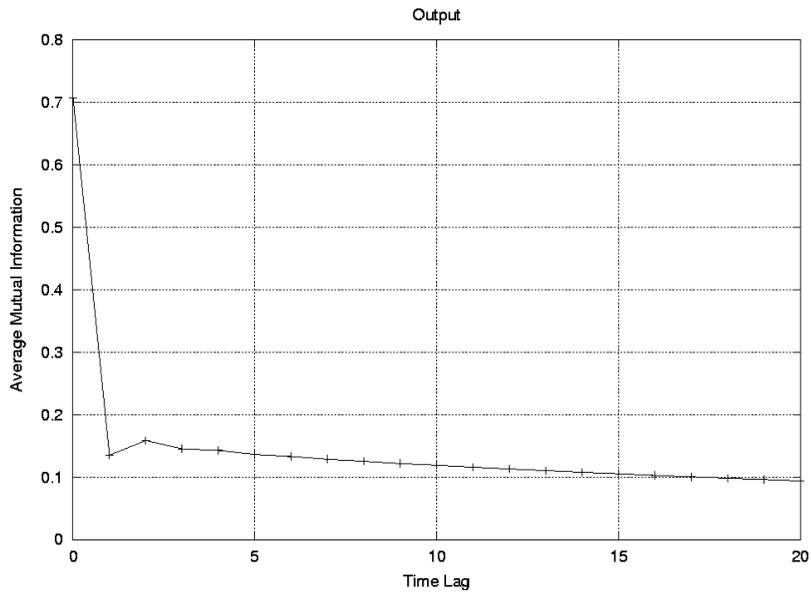}
\end{center}
\caption{Average mutual information.} 
\end{figure}

\begin{figure}
\begin{center}
\includegraphics[width=11cm,height=8cm]{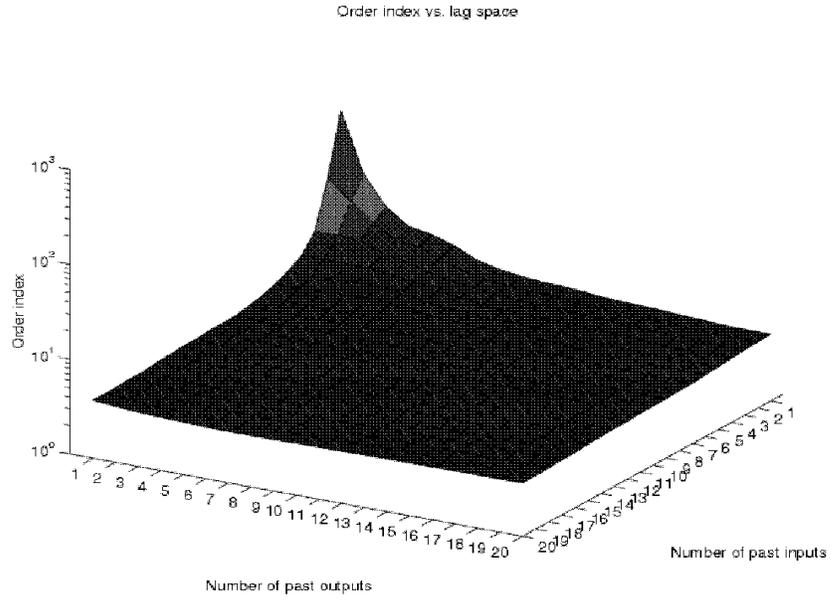}
\end{center}
\caption{Determination of the embedding dimension.}
\end{figure}

\begin{figure}
\begin{center}
\includegraphics[width=10cm,height=8cm]{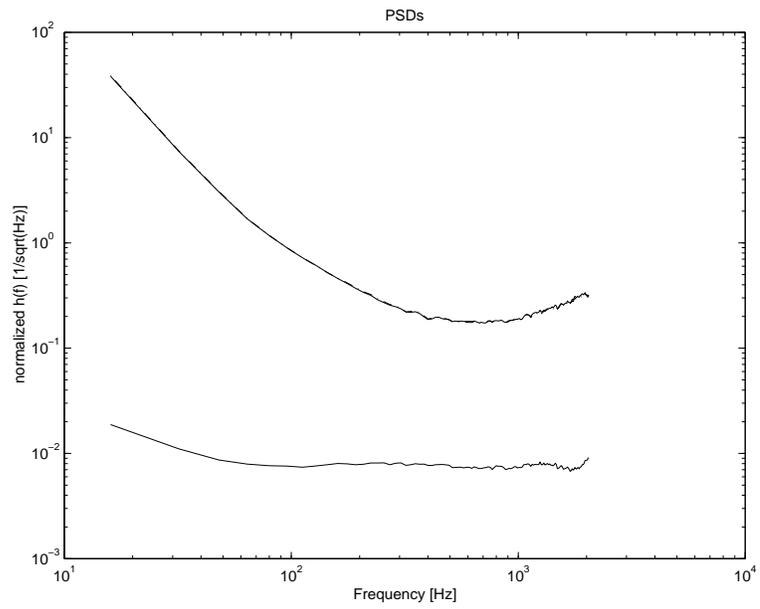}
\end{center}
\caption{Results of the training.}
\end{figure}

\begin{figure}
\begin{center}
\includegraphics[width=10cm,height=8cm]{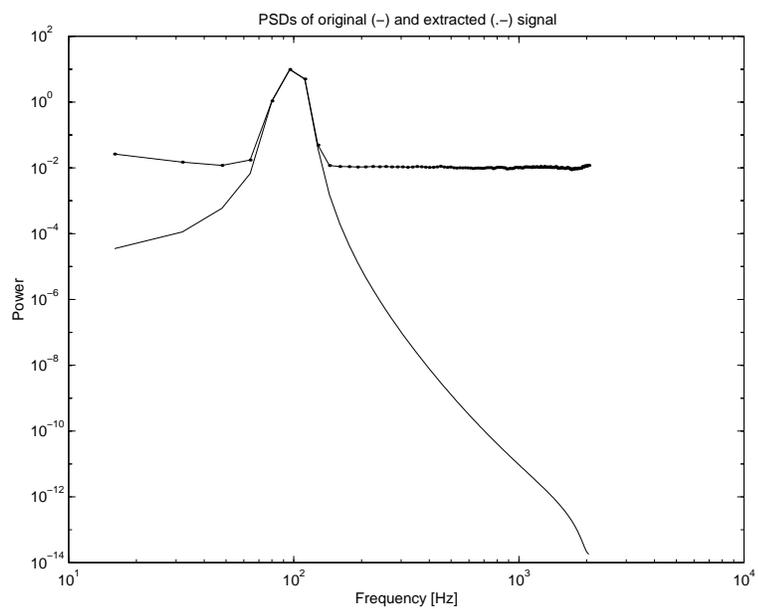}
\end{center}
\caption{Example of signal extraction.}
\end{figure}

\end{document}